# Tunable symmetry breaking in a hexagonal-stacked moiré magnet

Zeliang Sun*,1, Gaihua Ye*,2, Xiaohan Wan[1], Ning Mao[3], Cynthia Nnokwe[2], Senlei Li[4], Nishkarsh Agarwal[5], Siddhartha Sarkar[6], Zixin Zhai[7], Bing Lv[7], Robert Hovden[5], Chunhui Rita Du[4], Yang Zhang[8,9], Kai Sun[1], Rui He[+,2], Liuyan Zhao[+,1]

[1] Department of Physics, University of Michigan, Ann Arbor, MI, 48109, USA
[2] Department of Electrical and Computer Engineering, Texas Tech University, Lubbock, TX, 79409, USA
[3] Max Planck Institute for Chemical Physics of Solids, Nöthnitzer Straße 40, 01187 Dresden, Germany
[4] School of Physics, Georgia Institute of Technology, Atlanta, GA, 30332, USA
[5] Department of Materials Science and Engineering, University of Michigan, Ann Arbor, MI, 48109, USA
[6] Max Planck Institute for the Physics of Complex Systems, Nöthnitzer Straße 38, 01187 Dresden, Germany
[7] Department of Physics, the University of Texas at Dallas, Richardson, TX, 75089, USA
[8] Department of Physics and Astronomy, University of Tennessee, Knoxville, TN, 37996, USA
[9] Min H. Kao Department of Electrical Engineering and Computer Science, University of Tennessee, Knoxville, Tennessee 37996, USA

* The authors contribute equally
+ corresponding authors: Liuyan Zhao (lyzhao@umich.edu); Rui He (rui.he@ttu.edu)

**Abstract**

**Symmetry plays a central role in defining magnetic phases, making tunable symmetry breaking across magnetic transitions highly desirable for discovering non-trivial magnetism. Magnetic moiré superlattices, formed by twisting two-dimensional (2D) magnetic crystals, have been theoretically proposed and experimentally explored as platforms for unconventional magnetic states. However, despite recent advances, tuning symmetry breaking in moiré magnetism remains limited, as twisted 2D magnets, such as rhombohedral (R)-stacked twisted $CrI_3$, largely inherit the magnetic properties and symmetries of their constituent layers. Here, in hexagonal-stacked twisted double bilayer (H-tDB) $CrI_3$, we demonstrate clear symmetry evolution as the twist angle increases from 180º to 190º. While the net magnetization remains zero across this twist angle range, the magnetic phase breaks only the three-fold rotational symmetry at 180º, but it breaks all of the rotational, mirror, and time-reversal symmetries at intermediate twist angles between 181º and 185º, and all broken symmetries are recovered at 190º. These pronounced symmetry breakings at intermediate twist angles are accompanied by metamagnetic behaviors, evidenced by symmetric double hysteresis loops around zero magnetic field. Together, these results reveal that H-tDB $CrI_3$ at intermediate twist angles host a distinct moiré magnetic phase, featuring periodic in-plane spin textures with broken rotational, mirror, and time-reversal symmetries, which is markedly different from the out-of-plane layered antiferromagnetism in bilayer $CrI_3$ and the predominantly out-of-plane moiré magnetism in R-tDB $CrI_3$. Our work establishes H-stacked $CrI_3$ moiré magnets as a versatile platform for engineering magnetic properties, including and likely beyond complex spin textures.**

**Main Text**

Twisted two-dimensional (2D) magnets have demonstrated significant potential for realizing unconventional magnetic states[1-9] and enabling novel spintronic functionalities[10-12]. To date, the explored material platforms include twisted $CrI_3$ moiré superlattices, comprising constituent $CrI_3$ films ranging from monolayer[3] to bilayer[2,4,5,7] and trilayer[1,6,8], as well as twisted CrSBr "quasicrystals" with deliberately chosen large twist angles of 35° (ref. [11]) and 90° (ref. [9,10,12]), among others. However, the magnetic ground states observed in these twisted 2D magnets largely resemble those of their untwisted counterparts. Specifically, in twisted $CrI_3$, spins remain predominantly aligned along the out-of-plane direction[1,4,7,8], as in natural $CrI_3$ (ref. [13,14]), due to strong easy-axis anisotropy[15]. In twisted CrSBr, spins follow the easy axes of the individual constituent layers, attributed to weak interlayer coupling[16]. This substantial similarity between twisted and natural 2D magnets suggests that the influence of twist engineering has thus far been moderate, potentially limiting the realization of more exotic magnetic phenomena predicted for twisted 2D magnetic systems, such as skyrmion lattices[17-19], topological magnetism[20], and moiré or topological magnons[21,22]. These considerations motivate our investigation into the possibility of dramatically altering magnetic ground states through twist engineering, to the extent that significant symmetry breaking, absent in the natural layers, can be realized.

In this study, we select hexagonal-stacked twisted double bilayer (H-tDB) $CrI_3$, with twist angles close to 180°, as our sample platform. Due to the iodine octahedral cages surrounding the magnetic chromium sites, a $CrI_3$ monolayer exhibits only three-fold rotational symmetry ($C_3$), despite the honeycomb lattice of Cr atoms, rendering it inequivalent to its 180°-rotated counterpart[23] (Fig. 1a). Consequently, at the interface between stacked $CrI_3$ layers, the two nearest-neighbor iodine triangular facets are oriented oppositely in the near-0°, rhombohedral-stacked (R-stacked) configuration, but aligned in the same direction in the near-180°, H-stacked case (Fig. 1b). These contrasting interfacial geometries are expected to give rise to distinct lattice reconstructions and interlayer exchange couplings in R- and H-stacked twisted $CrI_3$ (ref. [15,24,25]). In our system of interest, H-tDB $CrI_3$ (Fig. 1c), each individual bilayer exhibits homogeneous monoclinic stacking, resulting in uniform antiferromagnetic (AFM) interlayer exchange coupling ($J_{inter}$). In contrast, the interface between the two bilayers forms a periodic moiré superlattice, leading to a spatially modulated interlayer exchange coupling ($J_{moiré}$). Within the moiré supercell of H-tDB $CrI_3$, three representative stacking sites are identified (Fig. 1d): H-AA, where Cr sites are directly aligned; H-AB, corresponding to rhombohedral stacking; and H-AB', associated with monoclinic stacking (Supplementary Note 1). First-principles calculations reveal a strong AFM interlayer coupling of 0.17 meV/$\mu_B^2$ at the H-AA site, an equally strong ferromagnetic (FM) interlayer coupling of –0.16 meV/$\mu_B^2$ at H-AB', and a weak AFM coupling of 0.04 meV/$\mu_B^2$ at H-AB (Supplementary Note 2). Furthermore, H-AB' and H-AC', that are mirror counterparts of each other, are identified as the structurally most favorable stacking site (Supplementary Note 3). These results stand in stark contrast to the R-stacked configuration, in which R-AB' and R-AB are structurally low-energy stacking sites, featuring strong FM interlayer coupling at R-AB, small AFM coupling at R-AB', and moderate FM coupling at R-AA[15].

We begin by characterizing the moiré superlattices of our hBN-encapsulated H-tDB $CrI_3$ samples using selected area electron diffraction (SAED) (see Methods). Figure 1e presents the SAED pattern from a sample with a targeted twist angle of 181.5°. The diffraction patterns from the H-tDB $CrI_3$ at the Miller indices $30\bar{3}0$, $33\bar{6}0$ and $60\bar{6}0$ reveal distinct moiré superlattice peaks arising from lattice reconstruction, in addition to the Bragg peaks of the two individual $CrI_3$ bilayers. The angular separation between the $CrI_3$ Bragg peaks is measured to be 181.54° ± 0.23°, closely matching the targeted twist angle of 181.5°. Compared with SAED patterns of R-tDB $CrI_3$ (ref. [2,4]), the moiré superlattice peaks in H-tDB $CrI_3$ are notably sharper and remain well-defined at higher Miller indices. This enhancement serves as compelling evidence of higher moiré superlattice quality in the H-stacked configuration, in agreement with theoretical predictions identifying H-AB' and H-AC' as the energetically preferred stacking site within the H-stacked moiré supercell (Supplementary Note 3).

We next investigate the magnetic properties of high-quality H-tDB $CrI_3$ samples. Figure 2a displays a magnetic circular dichroism (MCD) measurement result performed on a 181.1° H-tDB $CrI_3$ sample at 10 K under a variable out-of-plane magnetic field (see Methods). The MCD trace reveals a zero-magnetization ground state at zero magnetic field ($B = 0$ T), a spin-flip transition at a critical field of ± 0.7 T ($B_c = ± 0.7$ T), and a pronounced metamagnetic behavior characterized by symmetric magnetic hysteresis loops between $B_m = ±0.1$ T and $B_c$ (highlighted in red). Collectively, these features distinguish H-tDB $CrI_3$ from 1.1° R-tDB $CrI_3$ which possesses a finite magnetization at zero field[2,4] (Fig. 2b), and from natural bilayer $CrI_3$ which does not exhibit metamagnetic behaviors[13,26,27] (Fig. 2c). Temperature-dependent MCD measurements from 10 K to 60 K (Fig. 2d) further track the evolution of these features across the critical temperature ($T_c = 45$ K). Notably, the zero-field magnetization remains negligible throughout the entire temperature range (Fig. 2e), while the metamagnetic hysteresis (Fig. 2f) and transition field $B_m$ (Fig. 2g), *i.e.*, onset of metamagnetism, emerge just below $T_c$.

To assess the spatial uniformity of this zero-magnetization ground state in H-tDB $CrI_3$, we perform both scanning MCD microscopy and scanning nitrogen-vacancy (NV) magnetometry[1,8,28] on the 181.1° sample (see Methods). Across the 5 × 10 μm² twisted region outlined in the optical image of Fig. 2h, the MCD map shows a homogeneous zero-magnetization state at $B = 0$ T (Fig. 2i) and uniform saturated magnetization at $B = 2$ T (Fig. 2j), given the ~1 μm spatial resolution limited by optical diffraction. Complementary scanning NV magnetometry at $B = 0$ T, with a nanoscale ~70 nm resolution[8], further confirms the predominantly zero-magnetization state across the twisted area, with only a few localized regions of finite magnetization likely arising from fabrication-related imperfections (Fig. 2k). Together, the MCD and NV magnetometry measurements demonstrate the robustness and spatial uniformity of the zero-magnetization ground state in H-tDB CrI$_3$, in agreement with the high-quality moiré superlattices found in Figs. 1e and f.

To investigate the symmetry properties of this zero-magnetization ground state in H-tDB CrI$_3$, we perform polarization-resolved magneto-Raman spectroscopy in both linearly and circularly polarized bases (see Methods). This technique utilizes the magnetism-assisted phonon scattering processes to probe the symmetries of the underlying magnetic order, as previously demonstrated in natural few-layer CrI$_3$ (ref. [14,29-31]) and R-tDB CrI$_3$ (ref. [2,4]). We first present the Raman data obtained in the linear polarization basis. Figures 3a and 3b display the Raman spectra collected from a 183° H-tDB CrI$_3$ sample at 10 K in both parallel (upper panels) and crossed (lower panels) polarization channels, with incident linear polarizations ($\theta_{in}$) at 45° and 135°, respectively. In the parallel channel, a single Raman mode (U$_1$) appears with nearly identical intensities at both $\theta_{in}$ = 45° and 135°. In contrast, the crossed channel reveals four Raman modes (U$_1$, U$_2$, U$_3$, and U$_4$), attributed to Davydov splitting arising from the four-layer structure of H-tDB CrI$_3$. Notably, the intensity of U$_1$ (U$_2$) mode at $\theta_{in}$ = 45° is much larger (smaller) than that at $\theta_{in}$ = 135°, indicative of linear polarization anisotropy. This anisotropy contrasts with the isotropic behavior observed in R-tDB CrI$_3$ (ref. [2]). Furthermore, the U$_1$ mode in the parallel channel persists up to 80 K, whereas the U$_1$-U$_4$ modes in the crossed channel vanish at 80 K (Figs. 3c and d) and only emerge below the magnetic transition temperature $T_c$ = 45 K (Fig. 3e). This temperature dependence confirms that the U$_1$ mode in the parallel channel arises from conventional phonon scattering, while the U$_1$-U$_4$ modes in the crossed channel result from magnetism-assisted phonon scattering. Detailed polarization dependence of the U$_1$-U$_4$ modes in the crossed channel (Figs. 3f–i) exhibits a distinctive, rarely observed two-fold pattern[32-36]. Such a pattern requires Raman tensors with unequal real diagonal elements and antisymmetric real off-diagonal elements, of the form $\begin{pmatrix} a & c \\ -c & b \end{pmatrix}$ (Supplementary Note 4), which signifies the breaking of both C$_3$ rotational symmetry and vertical/diagonal mirror symmetries within the magnetic phase of H-tDB CrI$_3$.

We next examine the Raman response in the circular polarization basis. Figure 3j shows Raman spectra taken on the same 183° sample at 10 K in the co-circularly polarized RR and LL channels, where RR (LL) denotes right- (left-) handed circular polarization for both incident and scattered light. Three Raman modes (U$_1$, U$_3$, U$_4$) are resolvable in these two spectra, as the frequency difference between U$_1$ and U$_2$ is too small to separate them. Notably, a clear intensity difference exists between RR and LL channels at 10 K for all three modes, indicating Raman circular dichroism (CD), which disappears at 80 K (fig. 3k). Tracking the temperature dependence of the Raman CD, defined as $CD_{Raman} = (I_{LL}-I_{RR})/(I_{LL}+I_{RR})$, for modes U$_1$, U$_3$, and U$_4$ reveals its onset coinciding with the magnetic transition at $T_c$ = 45 K. Additionally, the magnetic field dependence of the Raman CD displays pronounced hysteresis loops below the spin-flip transition field $B_c$ (fig. 3m-o). The temperature and magnetic field dependencies of the Raman CD confirm its magnetic origin, requiring the Raman tensor to include antisymmetric imaginary off-diagonal elements, leading to the form $\begin{pmatrix} a & c+id \\ -c-id & b \end{pmatrix}$ (Supplementary Note 4) and hence reflecting the breaking of time-reversal symmetry in the magnetic phase of H-tDB CrI$_3$. The polarization dependence of the U$_1$-U$_4$ modes in the crossed channel (dots in Fig. 3f-i) can be well fitted by the Raman tensor $\begin{pmatrix} a & c+id \\ -c-id & b \end{pmatrix}$ (solid line in Fig. 3f-i).

Up to here, we have established the emergence of a zero-magnetization ground state that breaks C$_3$ rotational, vertical/diagonal mirror, and time-reversal symmetries in H-tDB CrI$_3$ at twist angles of 181.1° and 183°. To investigate the tunability of symmetry breaking, we performed MCD and magneto-Raman measurements at three additional twist angles, 180°, 185°, and 190°, as well as on a natural bilayer. Figures 4a–f present the MCD traces, while Figures 4g–l show the magneto-Raman spectra, all measured at 10 K. The 180° H-tDB CrI$_3$ consists of two antiparallelly aligned bilayers and serves as a reference without a moiré superlattice in the H-stacked case. Its MCD trace shows zero-magnetization at 0 T, very weak metamagnetic hysteresis loops, and two spin-flip transition (Fig. 4a), consistent with a layered AFM order similar to natural four-layer CrI$_3$ (ref. [37]). The Raman spectra are identical between $\theta_{in}$ = 45° and 135° for both parallel and crossed channels in the linear polarization basis, with no dichroism observed in the circular polarization basis (Fig. 4g). The detailed linear polarization dependence of U$_1$ Raman mode in the crossed channel follows a four-fold pattern (Fig. 4m), and, together with the absence of Raman CD, leads to a Raman tensor form of $\begin{pmatrix} a & 0 \\ 0 & b \end{pmatrix}$, suggesting only C$_3$ rotational symmetry breaking in 180° H-tDB CrI$_3$. H-tDB CrI$_3$ at intermediate twist angles, 181.1°, 183°, and 185°, exhibit similar behaviors, characterized by zero-magnetization at 0 T, metamagnetic hysteresis, and a single spin-flip transition in MCD (Figs. 4b-d), and Raman CD and a two-fold U$_1$ pattern in magneto-Raman data (Figs. 4h-j, n-p). Lastly in 190° H-tDB CrI$_3$, both MCD and Raman spectra converge to those of bilayer CrI$_3$ (Figs. 4e-f, k-i, q-r), featuring the absence of metamagnetic behavior and the presence of only two Raman modes. This suggests that the two bilayers in H-tDB CrI$_3$ decouple at such a large twist angle. We note that 190° H-tDB CrI$_3$ shows a Raman CD without the U$_1$ mode detected in the linearly crossed channel, pointing to a different origin from the broken time-reversal symmetry assigned at intermediate twist angle.

Figures 4s–v summarize the twist-angle dependence of the magnetic properties in H-tDB CrI$_3$. The metamagnetic behavior, characterized by the size of the hysteresis loop, displays a nonmonotonic trend with increasing twist angle, peaking at intermediate angles (Fig. 4s). Additionally, the breaking of C$_3$, mirror, and time-reversal symmetries – quantified for the U$_1$ mode by $|a–b|/|a+b|$, $|c|/|a+b|$, and $|d|/|a+b|$ (Fig. 4t-v),

respectively – follows a similar nonmonotonic pattern. The combination of metamagnetic behavior and the broken symmetries suggest the specialty of the magnetic ground state for the intermediate twist angle H-tDB CrI$_3$.

To further understand this magnetic ground state, we computationally mapped the magnetic phase diagram of H-tDB CrI$_3$ as a function of twist angle and magnetic anisotropy (see Methods and Supplementary Note 5). In the regime of low magnetic anisotropy and intermediate twist angles, we identified a magnetic ground state with modulated spin configurations in the two interfacial layers and homogeneous spins in the two outermost layers. In this state, the interfacial spins exhibit antiparallel in-plane alignment at H-AA site, and parallel out-of-plane alignment at H-AB′ and H-AB sites. Specifically, the spin texture across the four layers transitions from ↑←→↑ at H-AA to ↑↓↓↑ at H-AB′ and H-AB, resulting in a net out-of-plane magnetization of just 1% of the Cr$^{3+}$ magnetic moment per Cr atom averaged across the moiré supercell. This unique spin arrangement accounts for key experimental observations at intermediate twist angles: the spin sequence ↑←→↑ with in-plane components at H-AA sites breaks out-of-plane C$_3$ and vertical/diagonal mirror symmetries, while the minimal but finite out-of-plane magnetization breaks time-reversal symmetry but yields a negligible MCD signal. Notably, the emergence of in-plane spin components in this ground state requires a significantly reduced magnetic anisotropy in the two interfacial layers - decreased by 5-10 times of that in natural CrI$_3$ flakes and bulk[38]. We attribute this reduction to distortions of the iodine octahedral cage induced by moiré superlattice reconstruction.

In summary, we have identified a moiré magnetic phase in intermediate twist-angle H-tDB CrI$_3$ that exhibits strikingly different magnetic properties compared to the natural CrI$_3$ flakes and the R-stacked counterparts. These include the emergence of metamagnetism, the breaking of rotational, mirror, and time-reversal symmetries, and the development of a modulated spin texture, each and all of which confirm the profound impact of moiré engineering in H-tDB CrI$_3$. We attribute these much stronger moiré effects in H-tDB CrI$_3$ than R-tDB CrI$_3$ to the clearly higher structural quality of the moiré superlattices of the H-stacked system, as confirmed by their SAED patterns. Building on these findings, we anticipate the moiré engineering in H-tDB CrI$_3$ would make significant modifications to the magnon dispersions, which will further establish it as a promising platform for exploring moiré magnons.

## Methods

### Sample fabrication of H-tDB CrI$_3$

In this study, CrI$_3$ crystals were synthesized using the chemical vapor transport method, as described in previous literature[2,4]. The H-tDB CrI$_3$ devices were fabricated by the "tear-and-stack" method and were encapsulated by hBN flakes on both sides. To begin, we exfoliated the bulk CrI$_3$ and hBN crystals onto SiO$_2$/Si substrates to produce bilayer CrI$_3$ and few-layer hBN. A poly(bisphenol A carbonate) stamp was then used to sequentially pick up the top hBN and part of the bilayer CrI$_3$. The remaining bilayer CrI$_3$ on the Si/SiO$_2$ substrate was rotated to a controlled angle and then was subsequently picked up. The two bilayer CrI$_3$ flakes were stacked to create the twisted device, which was finally encapsulated with a bottom hBN flake. The encapsulated H-tDB CrI$_3$ for the Raman, MCD and NV measurement were transferred to the 285nm Si/SiO$_2$ substrate, and for the TEM measurement were transferred onto the TEM grid with a 10-nm-thick SiN membrane. All processes for fabricating the devices were conducted in a nitrogen-filled glovebox, where oxygen level was kept below 0.1 ppm and water level was kept below 0.5 ppm.

### TEM and SAED measurement

The Dark-Field (DF) imaging and Selected Area Electron Diffraction (SAED) of H-tDB CrI$_3$ were taken on a Thermo Fisher Talos instrument operated at 200 keV and equipped with a Gatan OneView camera. The SAED were acquired from a surveyed area of ~ 850 nm×850 nm. The twist angle was determined by fitting the $(03\bar{3}0)$, $(3\bar{3}00)$ and $(30\bar{3}0)$ Bragg peaks with two-dimensional Gaussian functions, and the average width of the fitted gaussians was used to estimate the error in the twist angle.

### Polarization-resolved magneto-Raman spectroscopy measurement

Polarized Raman spectra were taken using a Horiba LabRAM HR Evolution Raman microscope system with a thermoelectrically cooled charge-coupled device (CCD) camera. Backscattering geometry was employed in the Raman measurements with excitation laser with wavelength 633 nm. The laser beam was focused onto the sample with a full width at half maximum (FWHM) of 2–3 μm in diameter by using a ×40 transmissive objective. Temperature-dependent and magnetic field-dependent Raman spectra were conducted in a commercial variable-temperature, closed-cycle cryostat with a superconducting magnet (Cryo Industries of America, Inc.). For the determination of incident polarization direction relative to the sample axis, we first measured the polarization-dependent Raman spectra and fitted the intensity of U$_1$ mode in the crossed channel, then we defined the maximum of U$_1$ as the direction of 135°.

### Magnetic circular dichroism measurement

MCD measurement was conducted using excitation laser with wavelength of 633 nm. The incident light with linear polarization at angle 45° with respect to the PEM fast axis was modulated by a PEM with a retardance of λ/4. The reflected light was measured by a balanced amplified photodetector. The signal was demodulated by two lock-in amplifiers (Zurich Instruments, MFLI 500 kHz/5 MHz) separately, of which one lock-in amplifier was referenced to the PEM frequency 1f = 42.09 kHz and the other lock-in amplifier was referenced to the chopping frequency f = 511 Hz to get the total reflectance as a normalization. The MCD value was obtained by the normalization of 1f signal by the total reflectance. The MCD scanning measurements were performed in the attoDRY2100 with an attocube xyz piezo stage.

**NV magnetometry measurement**

The scanning NV magnetometry measurement were performed in a system consisting of a custom-designed atomic force microscope (AFM) operating in a cryo-free cryostat[8]. A diamond cantilever containing a single NV center was glued to a quartz tuning fork for force-feedback AFM operations, and a window on top of the cryostat provided optical access for NV measurements. The vertical distance between the NV sensor and the sample was set to be ~70 nm in the current study. An external microwave wire close to the diamond probe was used to apply microwave fields to control the NV spin states. We applied pulsed green laser and microwave signals to carry out NV optically detected magnetic resonance measurements. NV spin states were addressed by measuring NV photoluminescence using an avalanche photodiode. Scanning NV magnetometry exploits the Zeeman effect to quantitatively detect local magnetic stray fields (longitudinal to the NV spin axis) generated from H-tDB $CrI_3$ sample. Through well-established reverse-propagation protocols, the corresponding magnetization map of the sample can be reconstructed.

**First principles calculations**

We performed spin-polarized density functional theory (DFT) calculations using the Vienna Ab initio Simulation Package (VASP)[39], employing the projector-augmented wave method to treat core–valence interactions. A plane-wave cutoff energy of 450 eV was used, and the Perdew–Burke–Ernzerhof (PBE) functional within the generalized gradient approximation (GGA) was adopted for the exchange–correlation potential. Structural relaxations were carried out until the total energy and forces converged to $10^{-8}$ eV and 0.005 eV/Å, respectively. Brillouin-zone integrations were performed using a 9×9×1 Monkhorst–Pack k-point mesh. To account for electron correlation in the Cr $d$ orbitals, we employed the DFT+U method with an effective U of 3 eV. A vacuum region of more than 25 Å was introduced to avoid interactions between periodic images, and van der Waals corrections were included via the DFT-D2 method. To calculate the exchange parameter for each stacking configuration, we varied the shift vectors between the top and bottom layers from [0, 0] to [1, 1] and discretized into a 16×16 mesh in real space.

**Calculations of magnetic ground state of H-tDB CrI₃**

To determine the magnetic ground state of H-tDB CrI₃ at different twist angles, we numerically computed the energy of various spin configurations. Using numerical minimization, we identified the ground state configuration by finding the spin structure with the lowest total energy. The total Hamiltonian is given by:

$$H = H_1 + H_2 + H_3 + H_4 + H_{12} + H_{34} + H_{23}$$

where $H_1$, $H_2$, $H_3$ and $H_4$ describe the intralayer interactions for layers 1, 2, 3, and 4, respectively, while $H_{12}, H_{34}$, and $H_{23}$ represent the interlayer couplings between adjacent layers.

For the top and bottom layers, we define:

$$H_1 = H_4 = J_{\text{intra}} \sum_{<i,j>} (S_i^x S_j^x + S_i^y S_j^y + \gamma S_i^z S_j^z)$$

where $J_{\text{intra}}$ is the intralayer exchange coupling and $\gamma$ is the easy-axis magnetic anisotropy, with the values $J_{\text{intra}} = -2.2 \text{ meV}/\mu_B^2$ and $\gamma = 1.0445$ (ref. [4]).

For the middle two layers, we set:

$$H_2 = H_3 = J_{\text{intra}} \sum_{<i,j>} (S_i^x S_j^x + S_i^y S_j^y + \gamma_{\text{moiré}} S_i^z S_j^z)$$

where $J_{\text{intra}} = -2.2 \text{ meV}/\mu_B^2$, which is the same as layer 1 and 4, and $\gamma_{\text{moiré}}$ is a variable easy-axis magnetic anisotropy parameter, which denotes modified anisotropy of the middle two layers, to account for the effect of the distortion of iodine octahedral cages surrounding the magnetic chromium sites when forming moiré superlattice on the magnetic anisotropy of the two middle layers in H-tDB CrI₃.

For interlayer couplings, we set

$$H_{12} = J_{\text{inter}} \sum_l \mathbf{S}_l^{(1)} \cdot \mathbf{S}_l^{(2)},$$

$$H_{34} = J_{\text{inter}} \sum_l \mathbf{S}_l^{(3)} \cdot \mathbf{S}_l^{(4)},$$

where the coupling coefficient $J_{\text{inter}} = 0.04 \text{ meV}/\mu_B^2$ (ref. [4]). For the coupling between the middle two layers,

$$H_{23} = \sum_l J_{\text{moiré}} \mathbf{S}_l^{(2)} \cdot \mathbf{S}_l^{(3)},$$

where $J_{\text{moiré}}$ is obtained by fitting the first principles results up to six harmonics in the reciprocal lattice. The detailed calculation results and phase diagram are discussed in Supplementary Note 5.

## Data availability

All data supporting this work are available from the corresponding author upon request.

## Acknowledgements

L.Z. acknowledges support from the U.S. Department of Energy (DOE), Office of Science, Basic Energy Science (BES), under award No. DE-SC0024145. Z.S. acknowledges support from the National Science Foundation, under Grant No. DMR-2103731. R.He acknowledges support by NSF Grants No. DMR-2300640 and DMR-2104036 and DOE Office of Science Grant No. DE-SC0020334 subaward S6535A. K.S. and X.W. acknowledge the National Science Foundation through the Materials Research Science and Engineering Center at the University of Michigan, under award No. DMR-2309029. N.M. acknowledges the financial support from the Alexander von Humboldt Foundation. Y.Z. is supported by the startup fund at University of Tennessee. C.R.D. acknowledges support from the U.S. Department of Energy (DOE), Office of Science, Basic Energy Sciences (BES), under award No. DE-SC0024870 (for the hardware development of scanning NV microscopy) and Alfred. P. Sloan Foundation under award No. FG-2024-21387 (for the software development of scanning NV magnetometry). R.Hovden and N.A. acknowledge support from the National Science Foundation through the Materials Research Science and Engineering Center at the University of Michigan, Award No. DMR-2309029. B. L. and Z.Z. acknowledges the support from US Air Force Office of Scientific Research Grant No. FA9550-19-1-0037, National Science Foundation DMREF-2324033 and Office of Naval Research grant no. N00014-23-1-2020 and N00014-22-1-2755.

## Author contributions

Z.S., R.He and L.Z. conceived the idea and initiated this project. Z.S. fabricated natural $CrI_3$ and H-tDB $CrI_3$ samples. G.Y., Z.S. and C.N. performed magneto-Raman and MCD measurements. S.L. and C.R.D. performed scanning NV magnetometry measurements. N.A. and R.Hovden performed TEM measurements. Z.Z. and B.L. grew $CrI_3$ single crystals. M.N. and Y.Z. performed first principles calculations. X.W., S.S. and K.S. performed magnetic ground state calculations. Z.S., R.He and L.Z. analyzed the data and wrote the manuscript. All authors participated in discussions of the results.

## Competing Interests

The authors declare no competing interests.

**References**

1. Song, T. *et al.* Direct visualization of magnetic domains and moiré magnetism in twisted 2D magnets. *Science* **374**, 1140-1144 (2021).
2. Xie, H. *et al.* Twist engineering of the two-dimensional magnetism in double bilayer chromium triiodide homostructures. *Nature Physics* **18**, 30-36 (2022).
3. Xu, Y. *et al.* Coexisting ferromagnetic–antiferromagnetic state in twisted bilayer $CrI_3$. *Nature Nanotechnology* **17**, 143-147 (2022).
4. Xie, H. *et al.* Evidence of non-collinear spin texture in magnetic moiré superlattices. *Nature Physics* **19**, 1150-1155 (2023).
5. Cheng, G. *et al.* Electrically tunable moiré magnetism in twisted double bilayers of chromium triiodide. *Nature Electronics* **6**, 434-442 (2023).
6. Huang, M. *et al.* Revealing intrinsic domains and fluctuations of moiré magnetism by a wide-field quantum microscope. *Nature Communications* **14**, 5259 (2023).
7. Yang, B. *et al.* Macroscopic tunneling probe of Moiré spin textures in twisted $CrI_3$. *Nature Communications* **15**, 4982 (2024).
8. Li, S. *et al.* Observation of stacking engineered magnetic phase transitions within moiré supercells of twisted van der Waals magnets. *Nature Communications* **15**, 5712 (2024).
9. Healey, A. J. *et al.* Imaging magnetic switching in orthogonally twisted stacks of a van der Waals antiferromagnet. *arXiv preprint arXiv:2410.19209* (2024).
10. Boix-Constant, C. *et al.* Multistep magnetization switching in orthogonally twisted ferromagnetic monolayers. *Nature Materials* **23**, 212-218 (2024).
11. Chen, Y. *et al.* Twist-assisted all-antiferromagnetic tunnel junction in the atomic limit. *Nature* **632**, 1045-1051 (2024).
12. Boix‐Constant, C. *et al.* Programmable Magnetic Hysteresis in Orthogonally‐Twisted 2D CrSBr Magnets via Stacking Engineering. *Advanced Materials*, 2415774 (2025).
13. Huang, B. *et al.* Layer-dependent ferromagnetism in a van der Waals crystal down to the monolayer limit. *Nature* **546**, 270-273 (2017).
14. Jin, W. *et al.* Tunable layered-magnetism–assisted magneto-Raman effect in a two-dimensional magnet $CrI_3$. *Proceedings of the National Academy of Sciences* **117**, 24664-24669 (2020).
15. Sivadas, N., Okamoto, S., Xu, X., Fennie, C. J. & Xiao, D. Stacking-dependent magnetism in bilayer $CrI_3$. *Nano Letters* **18**, 7658-7664 (2018).
16. Liu, J., Zhang, X. & Lu, G. Moiré magnetism and moiré excitons in twisted CrSBr bilayers. *Proceedings of the National Academy of Sciences* **122**, e2413326121 (2025).
17. Tong, Q., Liu, F., Xiao, J. & Yao, W. Skyrmions in the Moiré of van der Waals 2D Magnets. *Nano Letters* **18**, 7194-7199 (2018).
18. Zheng, F. Magnetic Skyrmion Lattices in a Novel 2D‐Twisted Bilayer Magnet. *Advanced Functional Materials* **33**, 2206923 (2023).
19. Akram, M. *et al.* Moiré skyrmions and chiral magnetic phases in twisted $CrX_3$ (X= I, Br, and Cl) bilayers. *Nano Letters* **21**, 6633-6639 (2021).
20. Kim, K.-M., Kiem, D. H., Bednik, G., Han, M. J. & Park, M. J. Ab initio spin hamiltonian and topological noncentrosymmetric magnetism in twisted bilayer $CrI_3$. *Nano Letters* **23**, 6088-6094 (2023).
21. Wang, C., Gao, Y., Lv, H., Xu, X. & Xiao, D. Stacking domain wall magnons in twisted van der Waals magnets. *Physical Review Letters* **125**, 247201 (2020).
22. Li, Y.-H. & Cheng, R. Moiré magnons in twisted bilayer magnets with collinear order. *Physical Review B* **102**, 094404 (2020).

23  McGuire, M. A., Dixit, H., Cooper, V. R. & Sales, B. C. Coupling of crystal structure and magnetism in the layered, ferromagnetic insulator $CrI_3$. *Chemistry of Materials* **27**, 612-620 (2015).
24  Gibertini, M. Magnetism and stability of all primitive stacking patterns in bilayer chromium trihalides. *Journal of Physics D: Applied Physics* **54**, 064002 (2020).
25  Kong, X., Yoon, H., Han, M. J. & Liang, L. Switching interlayer magnetic order in bilayer $CrI_3$ by stacking reversal. *Nanoscale* **13**, 16172-16181 (2021).
26  Huang, B. *et al.* Electrical control of 2D magnetism in bilayer $CrI_3$. *Nature Nanotechnology* **13**, 544-548 (2018).
27  Jiang, S., Li, L., Wang, Z., Mak, K. F. & Shan, J. Controlling magnetism in 2D $CrI_3$ by electrostatic doping. *Nature Nanotechnology* **13**, 549-553 (2018).
28  Li, S. *et al.* Nanoscale magnetic domains in polycrystalline $Mn_3Sn$ films imaged by a scanning single-spin magnetometer. *Nano Letters* **23**, 5326-5333 (2023).
29  Huang, B. *et al.* Tuning inelastic light scattering via symmetry control in the two-dimensional magnet $CrI_3$. *Nature Nanotechnology* **15**, 212-216 (2020).
30  McCreary, A. *et al.* Distinct magneto-Raman signatures of spin-flip phase transitions in $CrI_3$. *Nature Communications* **11**, 3879 (2020).
31  Zhang, Y. *et al.* Magnetic order-induced polarization anomaly of Raman scattering in 2D magnet $CrI_3$. *Nano Letters* **20**, 729-734 (2019).
32  Hayes, W. & Loudon, R. Scattering of light by crystals.  (1978).
33  Kung, H.-H. *et al.* Chirality density wave of the "hidden order" phase in $URu_2Si_2$. *Science* **347**, 1339-1342 (2015).
34  Wang, Y. *et al.* Axial Higgs mode detected by quantum pathway interference in $RTe_3$. *Nature* **606**, 896-901 (2022).
35  Singh, B. *et al.* Uncovering the Hidden Ferroaxial Density Wave as the Origin of the Axial Higgs Mode in $RTe_3$. *arXiv preprint arXiv:2411.08322* (2024).
36  Cottam, M. G. & Lockwood, D. Light scattering in magnetic solids.  (1986).
37  Song, T. *et al.* Giant tunneling magnetoresistance in spin-filter van der Waals heterostructures. *Science* **360**, 1214-1218 (2018).
38  Kim, H. H. *et al.* Evolution of interlayer and intralayer magnetism in three atomically thin chromium trihalides. *Proceedings of the National Academy of Sciences* **116**, 11131-11136 (2019).
39  Kresse, G. & Furthmüller, J. Efficient iterative schemes for ab initio total-energy calculations using a plane-wave basis set. *Physical Review B* **54**, 11169 (1996).

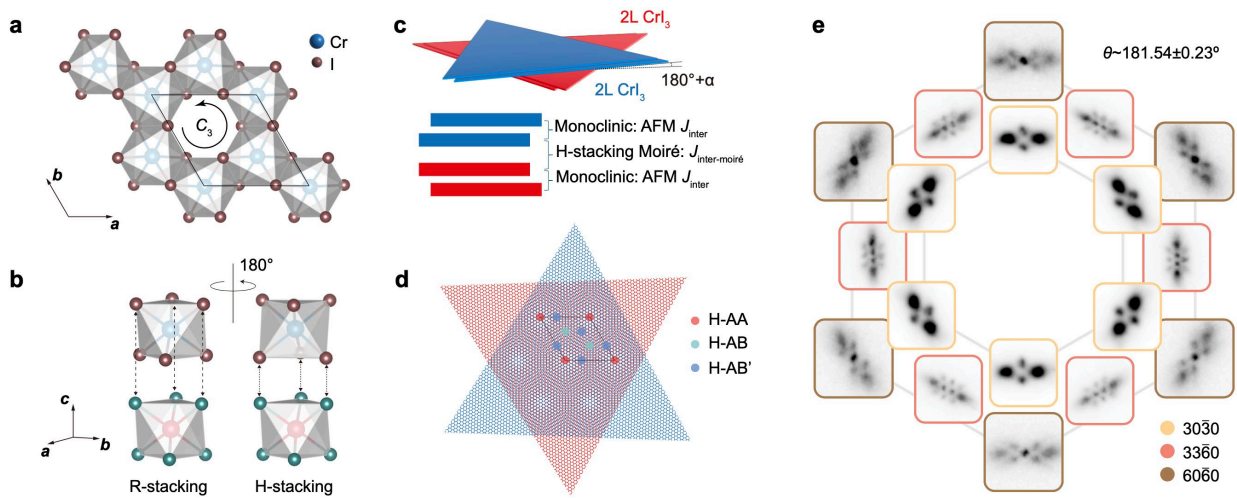

**Fig.1|Schematic structure of H-tDB CrI$_3$ and TEM results. a,** Crystal structure of monolayer CrI$_3$ with three-fold rotational symmetry. **b,** Schematic crystal structure of R-stacked bilayer CrI$_3$ (left) and H-stacked bilayer CrI$_3$ (right) at the AA site, respectively. The top layer of H-stacked bilayer CrI$_3$ is rotated by 180° relative to the R-stacked bilayer CrI$_3$. **c,** Schematic of H-tDB CrI$_3$ with twist angle 180°+α (upper panel). Periodically modulated interlayer exchange coupling ($J_{moiré}$) forms between the two 2L CrI$_3$, whereas individual 2L CrI$_3$ has monoclinic AFM $J_{inter}$ (lower panel). **d,** The moiré superlattice of H-tDB CrI$_3$ at the interface between two 2L CrI$_3$. Regions of H-AA (red), H-AB (green) and H-AB' (blue) stacking geometries are marked in one moiré supercell (black parallelogram). **e,** Zoomed-in Bragg peaks with the Miller indices $30\bar{3}0$, $33\bar{6}0$ and $60\bar{6}0$ of the SAED pattern of 181.54° H-tDB CrI$_3$.

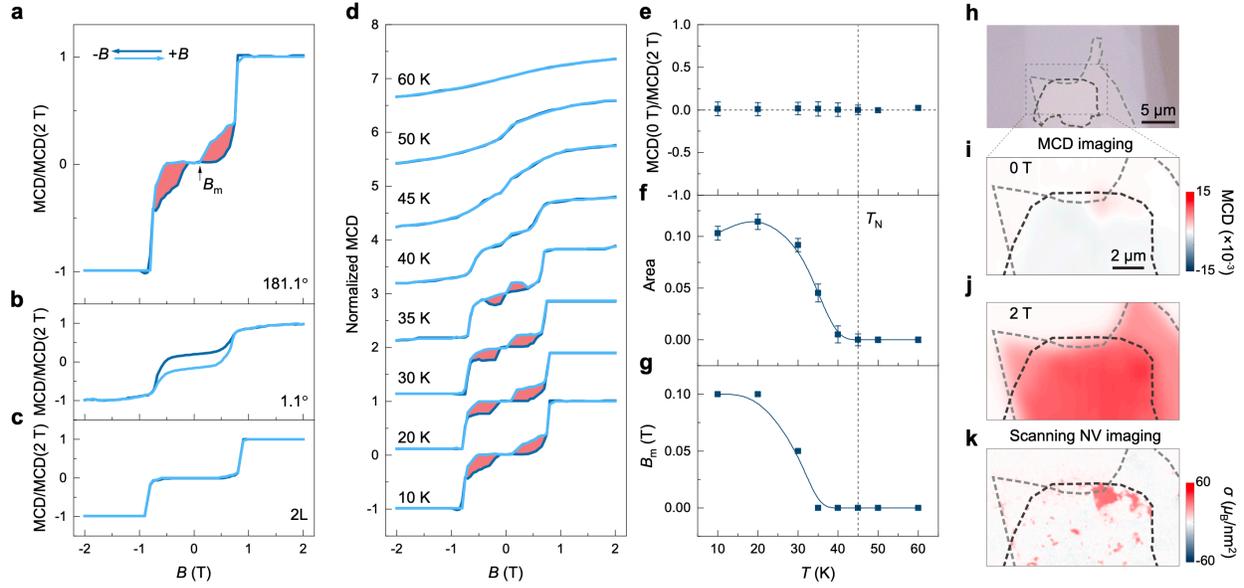

**Fig.2 | Magnetic field-dependent MCD data and NV magnetic image of 181.1° H-tDB CrI$_3$. a,** Magnetic field-dependent normalized MCD data of 181.1° H-tDB CrI$_3$ at $T$=10 K. The arrow marks the metamagnetic transition field $B_m$ and the red region denotes the magnetic hysteresis loop. The dark (light) blue curve represents the MCD data with decreasing (increasing) magnetic field. **b,** Magnetic field-dependent normalized MCD data of 1.1° R-tDB CrI$_3$ at $T$=10 K (data are taken from ref. [4]). **c,** Magnetic field-dependent normalized MCD data of 2L CrI$_3$ at $T$=10 K. **d,** Magnetic field-dependent MCD of 181.1° H-tDB CrI$_3$ at different temperatures. The MCD data at different temperatures are normalized by the value of MCD(2 T) at 10K and then shifted vertically. **e,** Temperature dependence of normalized MCD at 0T. **f,** Temperature dependence of the area of magnetic hysteresis loop (marked by red region in **d**). **g,** Temperature dependence of metamagnetic transition field $B_m$. **h,** Optical image of one 181.1° H-tDB CrI$_3$ sample. The region enclosed by the grey and black dashed lines is the twist area. **i,** MCD imaging at 0 T of the twist area marked in the rectangle in **h**. **j,** MCD imaging at 2 T of the twist area marked in the rectangle in **h**. **k,** Magnetization map obtained by scanning NV magnetometry measurements. The scanning area is the same as the MCD measurements with an out-of-plane magnetic field of 71 G applied. Images in **i-k** share the same scale bar shown in **i**.

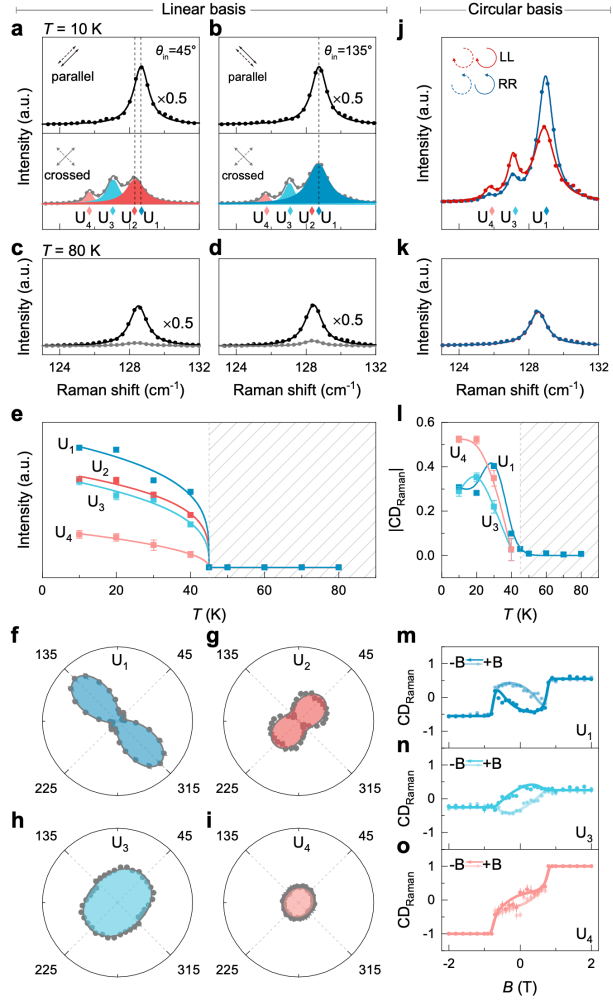

**Fig.3|Raman results of 183° H-tDB CrI$_3$. a, b,** Raman spectra of 183° H-tDB CrI$_3$ with **a,** $\theta_{in}$ =45° and **b,** 135° in both parallel (upper panel) and crossed (lower panel) channels at $T$=10 K, respectively. The colored Lorentzian profiles highlight the fitted individual U$_{1-4}$ modes in the crossed channel. The dash lines mark the position of U$_1$ and U$_2$ for comparison. **c, d,** Raman spectra of 183° H-tDB CrI$_3$ with **c,** $\theta_{in}$ =45° and **d,** 135° in both parallel (black line and dots) and crossed (grey line and dots) channels at $T$=80 K, respectively. **e,** Temperature-dependent fitted intensity of U$_{1-4}$ in crossed channel, in which the intensity of U$_1$ is fitted from the spectra with $\theta_{in}$ =135° and U$_{2-4}$ are fitted form the spectra with $\theta_{in}$ =45°. **f-i,** Polarization-dependent fitted intensity of U$_{1-4}$ in crossed channels at $T$=10 K, respectively. The solid lines are fitted curves by Raman tensor $\begin{pmatrix} a & c+id \\ -c-id & b \end{pmatrix}$. The intensity in **f-i** shares the same scale. **j,** Raman spectra of 183° H-tDB CrI$_3$ with circularly polarized light at $T$=10 K. Here, R (L) represents the right- (left-) handed circular polarization for incident light, which corresponds to the photon angular moment σ$_+$ (σ$_-$) with the positive (negative) sign defined along the upward (downward) out-of-plane direction. **k,** Raman spectra of 183° H-tDB CrI$_3$ with circularly polarized light at $T$=80 K. Only U$_1$ mode shows with the same intensity in LL and RR channels, U$_3$ and U$_4$ disappear above $T_N$. **l,** Temperature-dependent Raman circular dichroism |CD$_{Raman}$|=|($I_{LL}$-$I_{RR}$)/($I_{LL}$+$I_{RR}$)| of U$_1$, U$_3$ and U$_4$. **m-o,** Magnetic field-dependent Raman circular dichroism CD$_{Raman}$ of U$_1$, U$_3$ and U$_4$ at $T$=10 K, respectively.

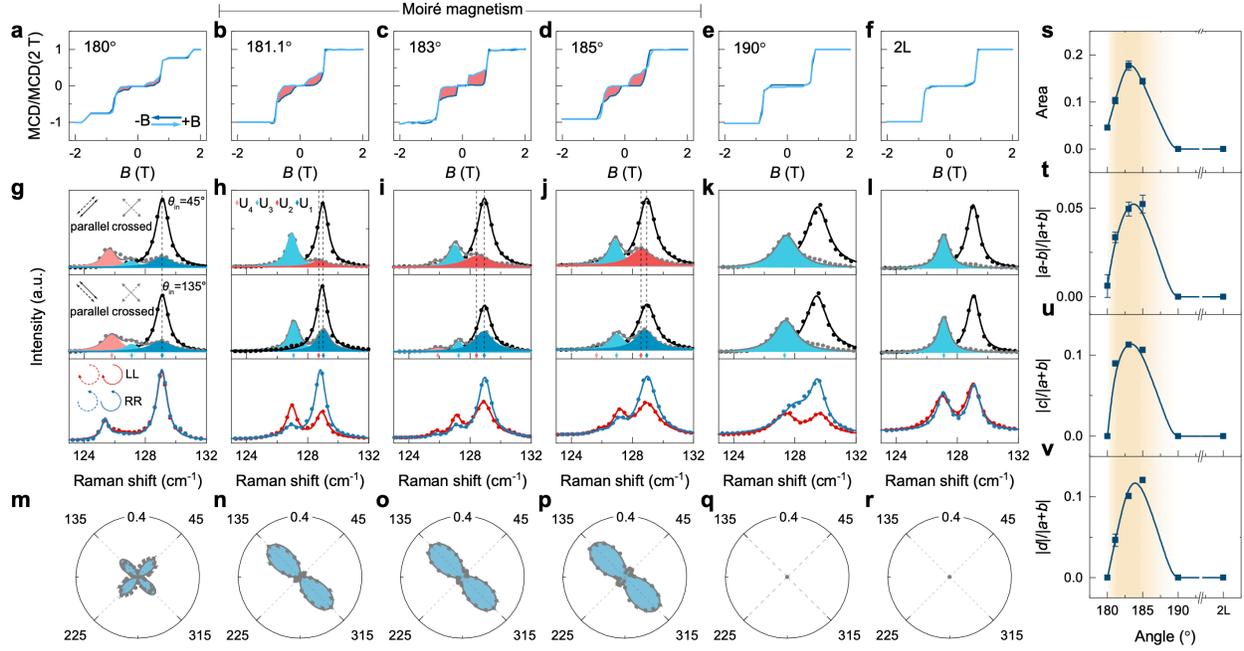

**Fig.4 | Twist angle dependence of MCD and Raman spectra of H-tDB CrI$_3$. a-f,** Normalized MCD data of H-tDB CrI$_3$ with twist angle from 180° to 190° and 2L CrI$_3$ taken at 10K. **g-l,** Raman spectra with linearly polarized light at incidence angle $\theta_{in}$=45° (upper panels), 135° (middle panels) and circularly polarized light (lower panels) at $T$=10 K. The black lines and dots represent Raman spectra in parallel channel and grey lines and dots represent Raman spectra in crossed channel, respectively. The colored Lorentzian profiles highlight the fitted individual U$_{1-4}$ modes in the crossed channel. The dash lines mark the position of U$_1$ and U$_2$ for comparison. **m-r,** fitting results of polarization-dependent Raman intensity of U$_1$ in crossed channel. U$_1$ does not appear in 190° H-tDB and 2L CrI$_3$ in crossed channel, so zero intensity is shown in **q** and **r**. The intensity of U$_1$ in the crossed channel is scaled by the intensity in the parallel channel in each twist angle. **s,** Twist angle-dependent magnetic hysteresis loop area, marked by red region in **a-f**. **t,** Twist angle-dependent fitted Raman tensor $|a-b|/|a+b|$ of U$_1$. **u,** Twist angle-dependent fitted Raman tensor $|c|/|a+b|$ of U$_1$. **v.** Twist angle-dependent fitted Raman tensor $|d|/|a+b|$ of U$_1$. The shaded area in **s-v** marks the range of twist angles which exhibit the moiré spin texture.

# Supplementary information for "Tunable symmetry breaking in a hexagonal-stacked moiré magnet"

Zeliang Sun*,1, Gaihua Ye*,2, Xiaohan Wan1, Ning Mao3, Cynthia Nnokwe2, Senlei Li4, Nishkarsh Agarwal5, Siddhartha Sarkar6, Zixin Zhai7, Bing Lv7, Robert Hovden5, Chunhui Rita Du4, Yang Zhang8,9, Kai Sun1, Rui He+,2, Liuyan Zhao+,1

[1] Department of Physics, University of Michigan, Ann Arbor, MI, 48109, USA

[2] Department of Electrical and Computer Engineering, Texas Tech University, Lubbock, TX, 79409, USA

[3] Max Planck Institute for Chemical Physics of Solids, Nöthnitzer Straße 40, 01187 Dresden, Germany

[4] School of Physics, Georgia Institute of Technology, Atlanta, GA, 30332, USA

[5] Department of Materials Science and Engineering, University of Michigan, Ann Arbor, MI, 48109, USA

[6] Max Planck Institute for the Physics of Complex Systems, Nöthnitzer Straße 38, 01187 Dresden, Germany

[7] Department of Physics, the University of Texas at Dallas, Richardson, TX, 75089, USA

[8] Department of Physics and Astronomy, University of Tennessee, Knoxville, TN, 37996, USA

[9] Min H. Kao Department of Electrical Engineering and Computer Science, University of Tennessee, Knoxville, Tennessee 37996, USA

* The authors contribute equally

+ corresponding authors: Liuyan Zhao (lyzhao@umich.edu); Rui He (rui.he@ttu.edu)

**Supplementary Information Note 1. Stacking configuration of rhombohedral-staked (R-stacked) and hexagonal-stacked (H-stacked) CrI$_3$.**

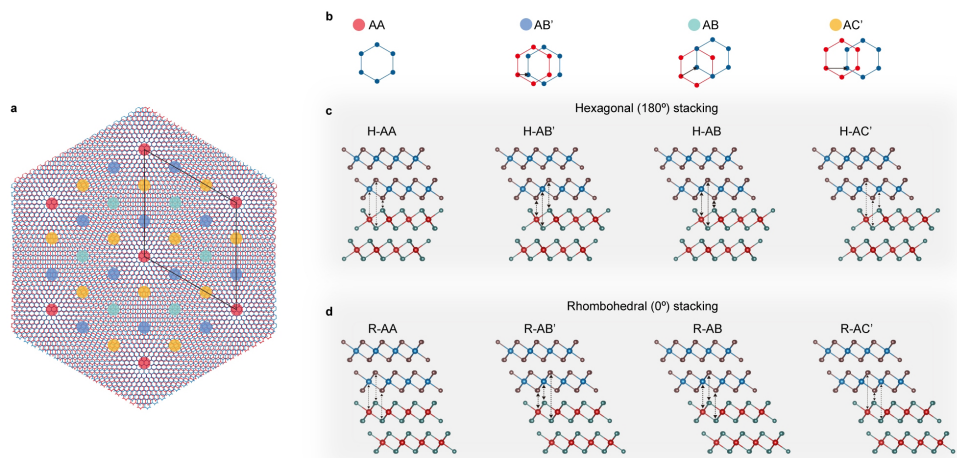

**Fig. S1| Stacking configuration of R-staked and H-stacked CrI$_3$. a,** The moiré superlattice of twisted double bilayer CrI$_3$ at the interface between two 2L CrI$_3$. Four representative stacking geometries are shown: AA (red dot), AB' (blue dot) AB (green dot) and AC' (yellow dot). **b,** Schematic structure of different stacking geometries shown in **a**. Here, only Cr atoms are shown, and the arrows indicate the shift vector between two CrI$_3$ layers. **c,** The side views of different stacking geometries of H-stacked double bilayer CrI$_3$. **d,** The side views of different stacking geometries of R-stacked double bilayer CrI$_3$.

**Supplementary Information Note 2. Calculated interlayer exchange coupling energy for H-stacked double bilayer CrI$_3$.**

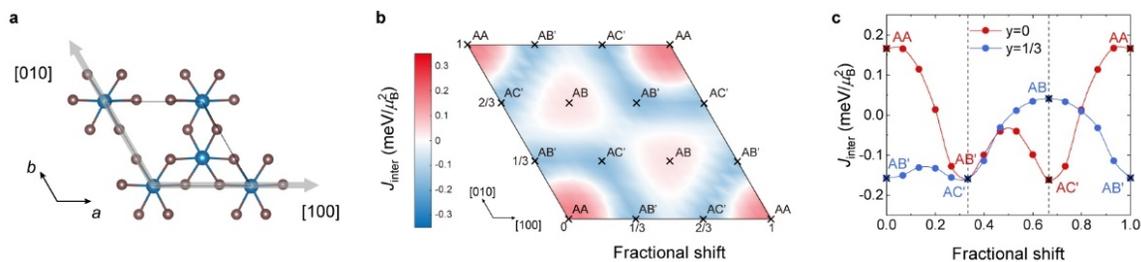

**Fig. S2|Interlayer exchange energy of H-stacked double bilayer CrI₃. a,** The crystal structure of monolayer CrI₃ in the *ab* plane. The shift vector [*x y*] defines the fractional coordinates of relative shift of the unit cell of one CrI₃ layer with respect to the unit cell in the neighboring layer of the interface, in which *x* is the fractional shift along [100] direction and *y* the fractional shift along [010] direction. For example, [0 0] is AA site, [1/3 0] is AB' site, [2/3 0] is AC' site, and [2/3 1/3] is AB site. **b,** Interlayer exchange energy of H-stacked double bilayer CrI₃ at different shift vectors. Four represent stacking sites AA, AB', AC' and AB are marked in the 2D map. **c,** The interlayer exchange energy along two high-symmetry line-cut [*x* 0] (red line) and [*x* 1/3] (blue line) of **b**.

| $J_{\text{inter}}$ (meV/$\mu_B^2$) | AA | AB | AB' | AC' |
|---|---|---|---|---|
| **R-stacking** | -0.102 | -0.314 | 0.039 | 0.063 |
| **H-stacking** | 0.166 | 0.040 | -0.158 | -0.162 |

**Table S1|Calculated interlayer exchange energy at four sites: AA, AB, AB' and AC' in R-stacked and H-stacked double bilayer CrI₃, respectively.**

**Supplementary Information Note 3. Calculated stacking energy for H-stacked double bilayer CrI₃.**

In this section, we discuss the formation of highly uniform moiré superlattice from the stacking energy in H-tDB CrI₃. Figure S3 shows the calculated stacking energy for H-stacked double bilayer CrI₃ at different stacking sites. It shows that the AA site has the largest stacking energy, then AB site, and the stacking energy of AB' and AC' are degenerate and lowest. In the moiré superlattice, the AA site is surrounded by the alternating degenerate AB' and AC' site (Fig. S1a), forming six-fold degenerate stacking energy distribution. Due to the more uniform distribution of stacking energy in the moiré superlattice, the H-tDB CrI₃ is expected to form higher-quality moiré superlattice.

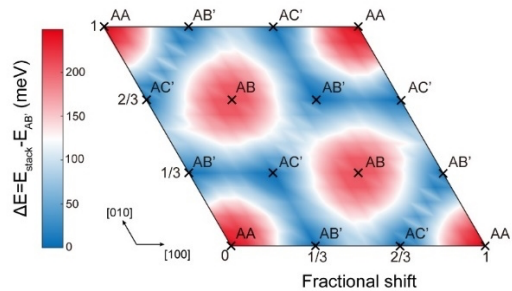

**Fig. S3|Calculated stacking energy for H-stacked double bilayer CrI₃ at different shift vectors.**

**Supplementary Information Note 4. Derivation of Raman tensor.**

This section shows our calculation of the Raman tensors of H-tDB CrI₃. As we used backscattering geometry with incident light perpendicular to the *ab* plane in the experiment, the Raman tensor *R* is a 2×2 matrix. The measured Raman intensities ($I$) can be expressed as: $I \propto |\langle \hat{e}_f | R | \hat{e}_i \rangle|^2$, where $\hat{e}_i$ and $\hat{e}_f$ are the unit polarization vectors of the incident and scattered light, respectively. In the parallel channel, in which $|\hat{e}_i\rangle$ is parallel to $|\hat{e}_f\rangle$, the incident and scattered light can be written as $|\hat{e}_i\rangle = \begin{pmatrix} cos\theta \\ sin\theta \end{pmatrix}$, and $|\hat{e}_f\rangle = \begin{pmatrix} cos\theta \\ sin\theta \end{pmatrix}$. In the crossed channel, in which $|\hat{e}_i\rangle$ is perpendicular to $|\hat{e}_f\rangle$, the incident and scattered light can be written as $|\hat{e}_i\rangle = \begin{pmatrix} cos\theta \\ sin\theta \end{pmatrix}$, and $|\hat{e}_f\rangle = \begin{pmatrix} -sin\theta \\ cos\theta \end{pmatrix}$. Here, we only show the derivation of Raman tensor and fitted intensity for $U_1$ mode, the fitted intensity of $U_2$-$U_4$ modes can be derived with the same Raman tensor.

First, in the AFM state of 2L $CrI_3$, only one peak $U_1$ shows in parallel channel with same intensity at different incident light angles (Table S2), consequently the Raman tensor in 2L $CrI_3$ for $U_1$ is $\begin{pmatrix} a & 0 \\ 0 & a \end{pmatrix}$. The intensity $I_{U_1,2L}^{parallel} \propto |(cos\theta, sin\theta) \begin{pmatrix} a & 0 \\ 0 & a \end{pmatrix} \begin{pmatrix} cos\theta \\ sin\theta \end{pmatrix}|^2 = a^2$, and $I_{U_1,2L}^{crossed} \propto |(-sin\theta, cos\theta) \begin{pmatrix} a & 0 \\ 0 & a \end{pmatrix} \begin{pmatrix} cos\theta \\ sin\theta \end{pmatrix}|^2 = 0$. We note that although the point group of 2L $CrI_3$ is $C_{2h}$, the Raman tensor of $U_1$ shows the $A_g$ symmetry of the $D_{3d}$ point group of monolayer $CrI_3$, which is attributed to the weak interlayer coupling. In 180°-tDB $CrI_3$, the pattern for $U_1$ in crossed channel shows four lobs with same intensity (fig. 4m in the main text), which breaks the rotational symmetry, and the Raman tensor is $\begin{pmatrix} a & 0 \\ 0 & b \end{pmatrix}$. Thus, the intensity $I_{U_1,180}^{parallel} \propto |(cos\theta, sin\theta) \begin{pmatrix} a & 0 \\ 0 & b \end{pmatrix} \begin{pmatrix} cos\theta \\ sin\theta \end{pmatrix}|^2 = |\frac{a+b}{2} + \frac{a-b}{2} cos2\theta|^2 = \frac{1}{8}(a-b)^2 cos4\theta + \frac{1}{2}(a^2-b^2)cos2\theta + \frac{1}{8}(3a^2 + 2ab + 3b^2)$, and $I_{U_1,180}^{crossed} \propto |(-sin\theta, cos\theta) \begin{pmatrix} a & 0 \\ 0 & b \end{pmatrix} \begin{pmatrix} cos\theta \\ -sin\theta \end{pmatrix}|^2 = |\frac{a-b}{2} sin2\theta|^2 = -\frac{1}{8}(a-b)^2 cos4\theta + \frac{1}{8}(a-b)^2$. For the twist angle of 181.1, 183 and 185, the patterns of $U_1$ in the crossed channel show four lobes with unequal intensity. Thus, the Raman tensor contains antisymmetric real off-diagonal Ramman tensor. Due to the nonzero Raman circular dichroism $I_{LL} \neq I_{RR}$, an antisymmetric complex off-diagonal Ramman tensor also appears in the Raman tensor. Consequently, the Raman tensor for $U_1$ is $\begin{pmatrix} a & c+di \\ -c-di & b \end{pmatrix}$. Here, we show the fitting process of 183° H-tDB $CrI_3$. The intensity $I_{U_1,183}^{parallel} \propto |(cos\theta, sin\theta) \begin{pmatrix} a & c+di \\ -c-di & b \end{pmatrix} \begin{pmatrix} cos\theta \\ sin\theta \end{pmatrix}|^2 = |\frac{a+b}{2} + \frac{a-b}{2} cos2\theta|^2 = \frac{1}{8}(a-b)^2 cos4\theta + \frac{1}{2}(a^2-b^2)cos2\theta + \frac{1}{8}(3a^2 + 2ab + 3b^2)$, and $I_{U_1,183}^{crossed} \propto |(-sin\theta, cos\theta) \begin{pmatrix} a & c+di \\ -c-di & b \end{pmatrix} \begin{pmatrix} cos\theta \\ sin\theta \end{pmatrix}|^2 = |c + id + \frac{a-b}{2} sin2\theta|^2 = -\frac{1}{8}(a-b)^2 cos4\theta + (a-b)csin2\theta + \frac{1}{8}(a-b)^2 + c^2 + d^2$. In the circular basis, for 183°-tDB $CrI_3$, $I_{U_1,183}^{LL} \propto |(1,-i) \begin{pmatrix} a & c+di \\ -c-di & b \end{pmatrix} \begin{pmatrix} 1 \\ i \end{pmatrix}|^2 = (a+b-2d)^2 + 4c^2$ and $I_{U_1,183}^{RR} \propto |(1,i) \begin{pmatrix} a & c+di \\ -c-di & b \end{pmatrix} \begin{pmatrix} 1 \\ -i \end{pmatrix}|^2 = (a+b+2d)^2 + 4c^2$. Because $I_{U_1,183}^{LL} \neq I_{U_1,183}^{RR}$, Raman circular

dichroism is shown. The same fitting analysis also applies to the 181.1° and 185°-tDB CrI$_3$ samples. We summarize the Raman tensors and polarization-dependent intensity pattern in Table S2.

| | Raman tensor | parallel channel | crossed channel | Raman circular dichroism |
|---|---|---|---|---|
| 2L | $\begin{pmatrix} a & 0 \\ 0 & a \end{pmatrix}$ | 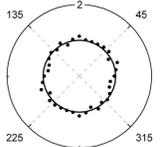 | | $I_{LL}=I_{RR}=4|a|^2$ |
| 180° | $\begin{pmatrix} a & 0 \\ 0 & b \end{pmatrix}$ | 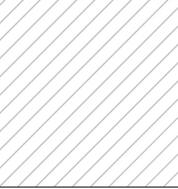 | 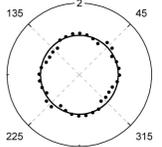 | $I_{LL}=I_{RR}=|a+b|^2$ |
| 183° | $\begin{pmatrix} a & c+id \\ -c-id & b \end{pmatrix}$ | 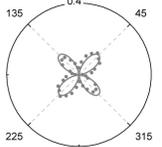 | 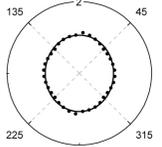 | $I_{LL}=(a+b-2d)^2+4c^2$ $I_{RR}=(a+b+2d)^2+4c^2$ $I_{LL}\neq I_{RR}$ |

**Table S2|Raman tensor of 180°, 183° H-tDB and 2L CrI$_3$.**

**Supplementary Information Note 5. Calculations of magnetic ground state of H-tDB CrI$_3$.**

In this section, we discuss the ground spin state of H-tDB CrI$_3$ at different twist angles. Using numerical minimization, we identified the ground spin state by finding the spin structure with the lowest total energy of the Hamiltonian given in the method section of the main text. To account for the effect of possible structural distortion of iodine octahedral cages caused by the moiré superlattice in tDB CrI$_3$, we varied the magnetic anisotropy $\gamma_{\text{moiré}}$ for the middle two layers, as well as for all the four layers, in the calculations.

In Figure S4a, we plot the $J_{\text{moiré}}$ function, fitted from the shift vector-dependent $J_{\text{inter}}$ as shown Fig. S2b. By minimizing the energy, we obtain the ground state spin configuration at different twist angles and $\gamma_{\text{moiré}}$ with a triangular grid of $L \times L \times 4 = 100 \times 100 \times 4$. The phase diagram of H-tDB CrI$_3$ with varied twist angle and magnetic anisotropy $\gamma_{\text{moiré}}$ is shown in Fig. S4c and S4d, where the horizontal axis is $\gamma_{\text{moiré}} - 1$ and vertical axis is the twist angle. The phase diagrams in Figs. S4c (varying $\gamma_{\text{moiré}}$ for the middle two layers at the interface) and S4d (varying $\gamma_{\text{moiré}}$ throughout all four layers) are qualitatively similar, featuring three prominent distinct phases (see schematic spin configurations of different phases in Fig. S4b), which we label as the twisted phase with one out-of-plane domain wall (O-1DW), the twisted phase with two in-plane domain walls (I-2DW), and the collinear phase (CL, no domain walls). At large twist angles, the middle two layers favor FM ground states, showing CL phase. The middle two layers favor FM ground states at large twist angles because the large area of AB' region with FM interlayer exchange coupling dominates.

Notably, for $1.00445 < \gamma_{moiré} < 1.00846$, the I-2DW phase appears over a broad range of twist angles. I-2DW phase features in-plane domains (major component of spins within the domain are in-plane) at H-AA. As shown in Fig. S4e, in the I-2DW phase ($1.00445 < \gamma_{moiré} < 1.00846$), the net magnetization $\langle m_z \rangle$ is always of order 0.01, with 1 representing the magnetic moment of $Cr^{3+}$. As a reference, the computed net $\langle m_z \rangle$ in 1.1° R-tDB $CrI_3$ is of order 0.1 (ref. [1]). When $\gamma_{moiré}$ further increases, the favored ground state is O-1DW phase, which is similar to the ground state of R-tDB $CrI_3$ with twist angle around 1.1° (ref. [1]). The net magnetization $\langle m_z \rangle$ in O-1DW phase also shows larger values compared to I-2DW phase. From the calculation results, the experimentally observed moiré spin textures in H-tDB $CrI_3$ with twist angle 181.1°~185° likely corresponds to the magnetic ground state of I-2DW phase, with in-plane domains and near zero net magnetization. The in-plane domains break $C_3$ rotational symmetry and mirror symmetry, and the allowed but very small net magnetization breaks the time-reversal symmetry but does not cause a large MCD signal. The combination of them is consistent with the form of Raman tensor obtained from experiment.

To compare, we also calculated the phase diagram considering all 4 layers with $\gamma_{moiré}$. We found that it is similar to the phase diagram considering only middle two layers with $\gamma_{moiré}$, except the ground states feature a 4-layer in-plane domain walls (I-4DW) between 181°~181.5° at small $\gamma_{moiré}$ and the ground state becomes twisted phase with 2 out-of-plane domain walls (O-2DW) around 180.8° (Fig. S4d). The calculated $<m_z>$ with $\gamma_{moiré}$ of all 4 layers are shown in Fig. s4f.

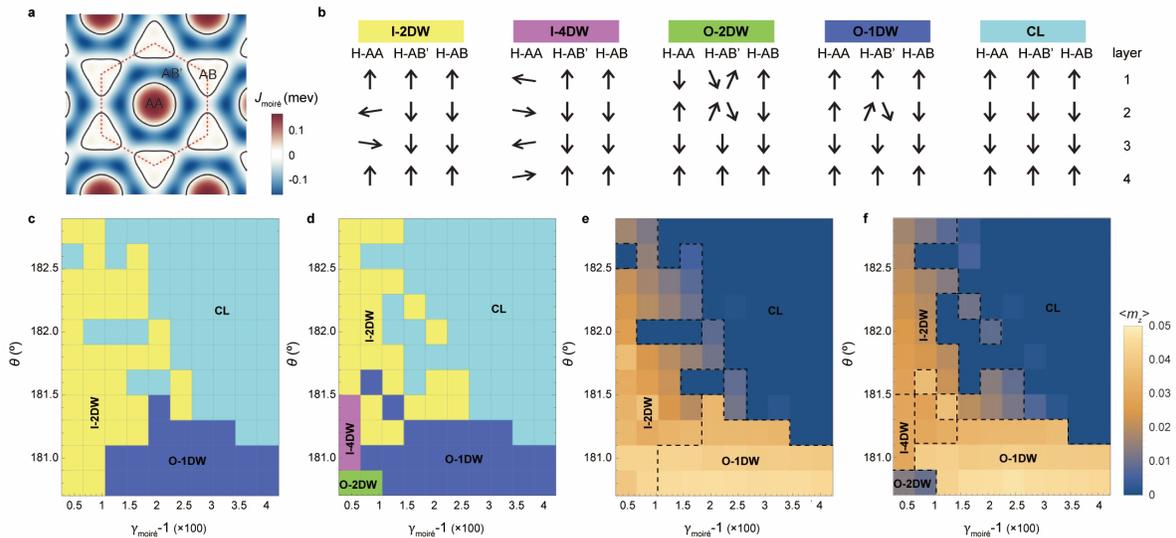

**Fig. S4|Calculation results of magnetic ground state of H-tDB CrI$_3$. a,** Fitted moiré interlayer coupling energy $J_{moiré}$. H-AA and H-AB have AFM coupling while H-AB' has FM coupling. The black lines label the place where $J_{moiré}=0$. **b,** Schematic spin configurations of different phases in **c** and **d**. **c,** Phase diagram of H-tDB CrI$_3$ with varied twist angles and magnetic anisotropy $\gamma_{moiré}$ of middle two layers. **d,** Phase diagram of H-tDB CrI$_3$ with varied twist angles and magnetic anisotropy $\gamma_{moiré}$ of all four layers. **e,** Net magnetization $<m_z>$ of H-tDB CrI$_3$ with varied twist angles and magnetic anisotropy $\gamma_{moiré}$ of middle two layers. **f,** Net magnetization $<m_z>$ of H-tDB CrI$_3$ with varied twist angles and magnetic anisotropy $\gamma_{moiré}$ of all four layers.

Having clarified the magnetic ground states at zero magnetic field, we further calculated the energy of different spin configurations and their out-of-plane magnetization ($m_z$) under magnetic field. First, we found that under magnetic field, the ground state I-2DW phase at zero field first transits to an intermediate phase with one out-of-plane domain walls with large $m_z$~0.43 (O-1DW-II, spin alignment is shown in Fig. S5). With further increasing field above 0.7 T, the ground state evolves to FM phase. Second, although the ground state at zero field is I-2DW phase, the ground state becomes O-1DW phase and the first excited phase becomes I-2DW phase when the field increases to 0.08 T. The energy difference between I-2DW and O-1DW are smaller than 0.1% between ±0.36 T. And $m_z$ of O-1DW phase under magnetic field is smaller than O-1DW-II phase. Based on the above calculation results, we can get the evolution of $m_z$ of H-tDB CrI$_3$ under magnetic field (Fig. S5). The initial ground state at zero magnetic field is I-2DW phase (yellow dot in Fig. S5) with near zero $m_z$. With increasing magnetic field (below 0.7 T), I-2DW phase first transits to O-1DW-II phase (red dot in Fig. S5) with a larger $m_z$ at $B_m$~0.1 T, which shows an abrupt upturn of $m_z$ curve above $B_m$. When the magnetic field is larger than 0.7, the spin state enters FM phase (cyan dot in Fig. S5). When decreasing the magnetic field below 0.7 T, the spin state first enters O-1DW phase (blue dot in Fig. S5) with smaller $m_z$ than O-1DW-II phase, thus forming a magnetization hysteresis loop as observed in the experimental data. As field further decreases below 0.1 T ($B_m$), the spin state re-enters the ground state (I-2DW).

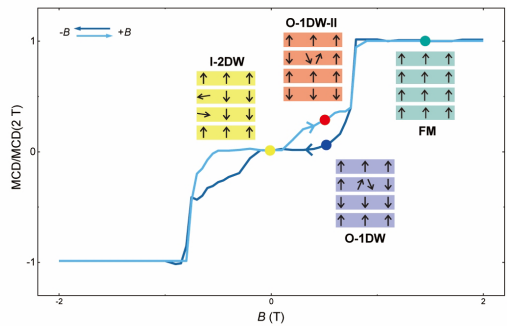

**Fig. S5|Evolustion of spin state of H-tDB CrI$_3$ under magnetic field.**

**Reference**

1. Xie, H. *et al.* Evidence of non-collinear spin texture in magnetic moiré superlattices. *Nat. Phys.* **19**, 1150-1155 (2023).